\documentclass[11pt,onecolumn]{article}
\usepackage{amsmath}
\usepackage{graphicx}
\usepackage{amssymb}
\usepackage{cite}
\newcommand{\dd}{\mathrm{d}}
\newcommand{\ii}{\mathrm{i}}
\newcommand{\ee}{\mathrm{e}}

\newcommand{\heav}{{\text{\usefont{U}{psy}{m}{n}\selectfont\symbol{113}}}}

\newcommand{\jell}{\mathrm{jell}}

\newcommand{\bulk}{\mathrm{bulk}}
\newcommand{\surf}{\mathrm{surf}}
\newcommand{\slab}{\mathrm{slab}}

\begin{document}




\title{The chemical potential and the work function of a metal film on a dielectric substrate}


\author{Kostrobij P. P., Markovych B. M.\\ \it Lviv Polytechnic National University   \\ \it             12 S. Bandera str., 79013, Lviv, Ukraine}
\maketitle

\begin{abstract}
 The chemical potential and the work function of an aluminum film,
 which (1) is in vacuum and (2) is located on a dielectric substrate is calculated within the model of non-interacting electrons located in an asymmetric rectangular potential well.
 For the first time in calculating these values for such a model of a metal film,
 the electroneutrality condition is correctly taken into account.
 This leads to the correct behavior of these values,
 namely:
 if the thickness of the film increases,
 these characteristics tend to their bulk values.
 
\end{abstract}




 \section{Introduction}

 Thin metal films on dielectric or semiconductor substrates exhibit properties that are interesting both from the point of view of fundamental science and from the perspectives of their technical applications in nano-sized electronic devices.

 One of methods for theoretical study of thin metal films is the use of model potentials,
 which are simple enough to solve the stationary Schr\"{o}dinger equation analytically,
 and secondly,
 to qualitatively correctly reflect the physical picture,
 namely,
 they do not allow electrons to leave a metal film.
 The simplest model potential is the infinite rectangular potential well,
 which,
 in particular,
 was used in Refs.~\cite{Thompson19636,Paskin1965A1965,Biao2008035410,Dymnikov2011901,Schulte1977149,Yong2009155404,Kostrobij201651,Kostrobij2017arxiv}.
 In Refs.~\cite{Thompson19636,Paskin1965A1965,Biao2008035410,Dymnikov2011901}
 (in fact,
 article~\cite{Dymnikov2011901} is a repetition of Ref.~\cite{Thompson19636} without reference to it),
 the chemical potential of non-interacting electrons in a film is calculated with the electroneutrality condition taken into account incorrectly.
 Authors of these articles incorrectly assumed that the film thickness equals to the width of the rectangular potential well.
 In fact,
 the film thickness is less than the width of the rectangular potential well by some distance,
 which was calculated in Ref.~\cite{Himbergen19782674}.
 This distance depends both on the width and height of the barrier of the rectangular potential well.
 For the first time, this distance was calculated by Bardeen~\cite{Bardeen1936653} for the semi-infinite jellium with the finite rectangular potential barrier.
 The need to take this distance into account was discussed in Refs.~\cite{Huntington19511035,Stratton1965556}.
 Due to this mistake made in Refs.~\cite{Thompson19636,Paskin1965A1965,Biao2008035410,Dymnikov2011901},
 the received values of the chemical potential were too big and with increasing the film thickness they do not approach to the bulk value of the chemical potential.
 Surprisingly,
 neither authors of the recent false article~\cite{Biao2008035410} in Phys. Rev. B,
 nor referees who recommended this article to be published did not know either about the article~\cite{Himbergen19782674},
 published in 1978 in the same journal,
 or about the already classical articles of Bardeen~\cite{Bardeen1936653} and Huntington~\cite{Huntington19511035} in the same journal.
 The same situation with another article~\cite{Pogosov2005195410} in the same journal,
 in which the work function and the electronic elastic forces for a film in the model of non-interacting electrons are calculated by using the finite rectangular potential well with the electroneutrality condition taken into account incorrectly.
 As a result,
 the calculated values of the work function are too small and with increasing the film thickness do not tend to its bulk value.
 The same mistake is made in Ref.~\cite{Kurbatsky2004543},
 in which the work function and the electronic elastic forces for the film is also calculated, and in the recent article~\cite{Korotun2015391},
 in which the work function for films,
 which are in the vacuum and on the dielectric substrate.
 Authors of Ref.~\cite{Kurbatsky2004543} came to the false conclusion that the work function of a low-dimensional metal structure is always less than the work function of a semi-infinite metal.

 The situation with articles~\cite{Schulte1977149,Yong2009155404} is somewhat better,
 in which for the model of non-interacting electrons the distance between the edge of the semi-infinite jellium and the infinite potential barrier was taken, instead of the correct distance between the edge of the film and the infinite potential barrier.
 That is,
 instead of the distance~\cite{Himbergen19782674},
 its value is used,
 when the film thickness approaches infinity.
 Surprisingly,
 such partially correct electroneutrality condition consideration in the model of non-interacting electrons has led to a very good agreement with the calculations of the chemical potential using the density functional theory within the local density approximation~\cite{Schulte1976427}.

 The aim of the presented work is to demonstrate that even the simple model of non-interacting electrons,
 which are situated in the asymmetric rectangular potential well,
 and the correct taking into account the electroneutrality condition give physically correct results for the chemical potential and the work function,
 which approach their bulk values with increasing the ﬁlm thickness.
 The chemical potential and the work function are calculated for an aluminum ﬁlm,
 which (1) is in the vacuum and (2) is located on a dielectric substrate.


 \section{Model}

\begin{figure}[hbtp]
  \centering
  \includegraphics[width=0.7\textwidth]{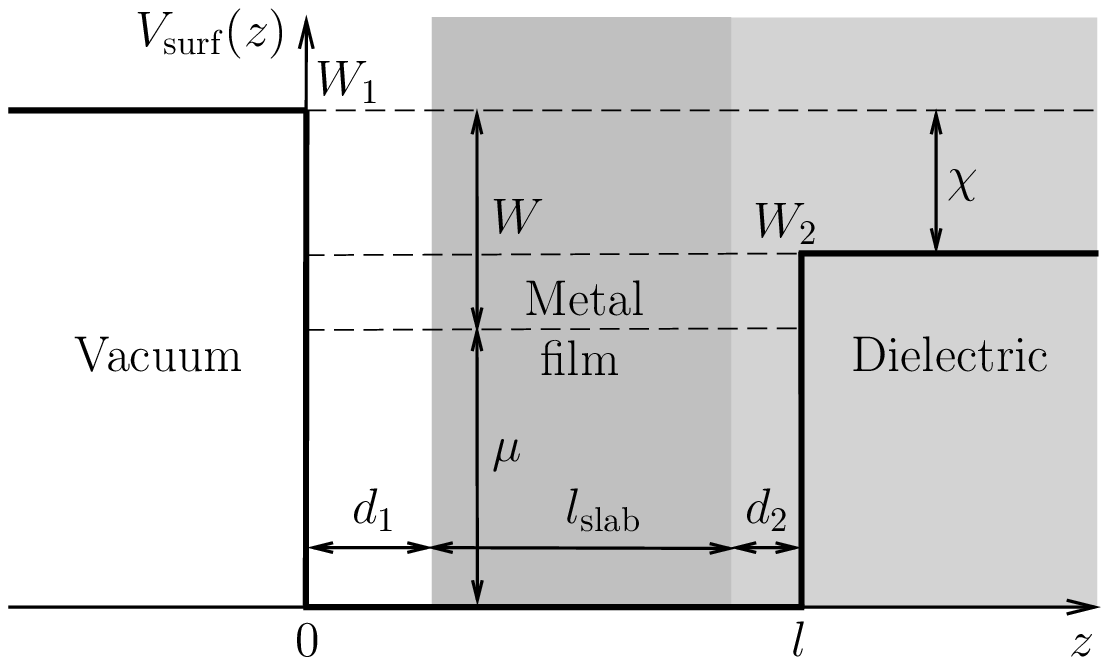}\\
  \caption{}\label{Slab}
\end{figure}

 We consider a metal film placed in such way that its two parallel infinite sides are parallel to the $xOy$ plane.
 Thickness of the slab is denoted by $l_\slab$ and lies along the $z$ axis.
 One side of the slab is specified by the equation ${z=d_1}$,
 and the second one is described by the equation ${z=l_\slab+d_1}$.
 An insulator is located to the right of the film,
 the electron affinity of which is $\chi$,
 and vacuum is to the left of the film as shown in Fig.~\ref{Slab}.

 The film is considered within the jellium model,
 i.e. an ionic subsystem is replaced by positive charge
 with the distribution
 \begin{align*}
   \varrho_\jell(\mathbf{r}_{||},z)
   &\equiv
   \varrho_\jell(z)
   =\varrho_0\,\heav(z-d_1)\,\heav(l_\slab+d_2-z)=\\
   &=\left\{
      \begin{array}{ll}
        \varrho_0, & z\in[d,l_\slab+d], \\
        0, & z\not\in[d,l_\slab+d],
      \end{array}
    \right.
 \end{align*}
 where $\heav(x)$ is the Heaviside step function,
 ${\textbf{r}_{||}=(x,y)}$,
 ${x,y\in(-\sqrt{S}/2,+\sqrt{S}/2)}$,
 ${z\in(-\infty,+\infty)}$,
 $S$ is area of the side of the film (${S\to\infty}$).
 The condition of electroneutrality is satisfied,
 \begin{equation}\label{electroNeutr}
  \lim_{S\to\infty}
  \int_S  \dd{\bf r}_{||}
  \int_{-\infty}^{+\infty}  \dd z \,
  \varrho_\jell({\bf r}_{||},z) = e N, \; e>0
 \end{equation}
 and withal,
 in the thermodynamic limit,
 we have
 \begin{equation*}
    \lim_{N,S\to\infty}\frac{eN}{Sl_\slab}=\varrho_0,
 \end{equation*}
 where $N$ is the number of electrons,
 which are situated in the field of the positive charge.

 As a consequence of the symmetry of the model,
 the motion of the electron in a plane parallel to the $xOy$ plane is free,
 and the one along the $z$ axis is determined by the surface potential $V_\surf(z)$.
 This potential is modeled by the asymmetric rectangular potential well with width $l$,
 namely,
 \begin{equation}\label{Vsurf}
   V_\surf(z)
   =\left\{
      \begin{array}{cc}
        W_1, & z\leqslant0, \\
        0,   & 0< z< l, \\
        W_2, & z\geqslant l.
      \end{array}
    \right.
 \end{equation}
 This model potential allows an analytical solving of the Schr\"odinger stationary equation,
 \begin{equation}\label{SchredEq}
   \left[
    -\frac{\hbar^2}{2m}\Delta+V_\surf(z)
   \right]\Psi_a(\mathbf{r})
   =E_a\Psi_a(\mathbf{r}),\;
   \mathbf{r}=(\mathbf{r}_{||},z)
 \end{equation}
 with the Dirichlet boundary conditions,
 \[
  \lim_{z\to\pm\infty}\Psi_a(\mathbf{r})=0,
 \]
 where $m$ is the electron mass,
 ${a=(\mathbf{k}_{||},\alpha)}$,
 $\mathbf{k}_{||}$ is  the two-dimensional wave vector of the electron in the plane parallel to the $xOy$ plane,
 $\alpha$ is a quantum number that characterizes the electron motion along the normal to the $xOy$ plane.

 We consider the bound states of electrons, i.e.,
 we suppose that the electron energy is not bigger than the smaller height of the potential walls,
 ${E_a\leqslant\min(W_1,W_2)}$.
 Then,
 from the conditions of continuity and smoothness of wave functions,
 we obtain an algebraic transcendental equation for the quantum number $\alpha$,
 \begin{equation}\label{eqForAlfa}
   \alpha l=\pi n-\arcsin\frac{\alpha}{s_1}-\arcsin\frac{\alpha}{s_2},\;
   n=1,2,\ldots,n_{\max},
 \end{equation}
 where ${s_i=\sqrt{2mW_i}/\hbar}$
 (${i=1,2}$),
 $n_{\max}$ is the number of bound states,
 \begin{multline}\label{nmax}
  n_{\max}=
  \left[
   \frac1\pi
   \left(
    l\min(s_1,s_2)+\arcsin\frac{\min(s_1,s_2)}{s_1}  \right.\right.\\
   \left.\left. +\arcsin\frac{\min(s_1,s_2)}{s_2}
   \right)
  \right],
  \end{multline}
 where square brackets denote taking the integer part.

 The wave function $\Psi_a(\mathbf{r})$ and the corresponding energy level $E_a$ of the electron in the bound state ${a=(\mathbf{k}_{||},\alpha)}$ are
 \[
  \Psi_a(\mathbf{r})
  =
  \frac1{\sqrt S}\ee^{\ii \mathbf{k}_{||} \mathbf{r}_{||}}
  \varphi_\alpha(z),\quad
  E_a=\frac{\hbar^2(k_{||}^2+\alpha^2)}{2m},
 \]
 \begin{align}\label{wavef}
    \varphi_\alpha(z)&=C(\alpha)\left\{
    \begin{array}{l}
      \displaystyle\tfrac{\alpha}{s_1}\ee^{\sqrt{s_1^2-\alpha^2}z}, \; z\leqslant0,\\
      \displaystyle\sin\left(\alpha z+\arcsin\tfrac{\alpha}{s_1}\right), \; 0<z< l,\\
      \displaystyle\sin\left(\alpha l+\arcsin\tfrac{\alpha}{s_1}\right)\ee^{-\sqrt{s_2^2-\alpha^2}(z-l)}, \; z\geqslant l,\\
    \end{array}
   \right.
 \end{align}
 where $C(\alpha)$ is the normalizing constant,
 \[
  C(\alpha)=\frac{\sqrt2}{\sqrt{l+\frac{(\alpha/s_1)^2}{\sqrt{s_1^2-\alpha^2}}+\frac{(\alpha/s_2)^2}{\sqrt{s_2^2-\alpha^2}}-
  \frac{\sin(\alpha l)\cos(\alpha l+2\arcsin\frac{\alpha}{s_1})}{\alpha}}}
 \]

 As can be seen from Fig.~\ref{Slab},
 there exists the following relation between the width of the potential well $l$ and the film thickness $l_\slab$,
 \begin{equation}\label{relatForllslab}
  l=l_\slab+d_1+d_2,
 \end{equation}
 where the parameters $d_i$ (${i=1,2}$) depend on the width of the potential well $l$ and are as follows~\cite{Himbergen19782674},
 \begin{equation}\label{d}
   d_i=\frac{3\pi}{8\mathcal{K}_\mathrm{F}}+\frac{\pi^2}{8\mathcal{K}_\mathrm{F}^2l} 
       -\frac{3}{4\mathcal{K}_\mathrm{F}}\left(\sqrt{\frac{s_i^2}{\mathcal{K}_\mathrm{F}^2}-1}
       +\left(2-\frac{s_i^2}{\mathcal{K}_\mathrm{F}^2}\right)\arcsin\frac{\mathcal{K}_\mathrm{F}}{s_i}\right),
 \end{equation}
 ${\mathcal{K}_\mathrm{F}=\sqrt{2m\mu}/\hbar}$ is the magnitude of the Fermi wave vector,
 $\mu$ is the chemical potential.

 It should be noted that if the heights of the potential barriers approach infinity,
 the parameter $d_i$ approaches the well-known magnitude ${\frac{3\pi}{8\mathcal{K}_\mathrm{F}}+\frac{\pi^2}{8\mathcal{K}_\mathrm{F}^2l}}$,
 which is the distance from the edge of the positive charge of the film to the potential wall~\cite{Kostrobij2017arxiv}.
 And if the width of the potential well $l$ approaches infinity,
 the parameter $d_i$ approaches the well-known magnitude
 ${\frac{3\pi}{8\mathcal{K}_\mathrm{F}}-\frac{3}{4\mathcal{K}_\mathrm{F}}\left(\sqrt{\frac{s_i^2}{\mathcal{K}_\mathrm{F}^2}-1}
       +\left(2-\frac{s_i^2}{\mathcal{K}_\mathrm{F}^2}\right)\arcsin\frac{\mathcal{K}_\mathrm{F}}{s_i}\right)}$,
 which is the distance from the edge of the positive charge to the potential well for the semi-infinite jellium
 (see, for example, \cite{Bardeen1936653,Huntington19511035,Kostrobij2015075441}).

 The chemical potential of $\mu$ of the electron gas must be found from the condition of electroneutrality~\eqref{electroNeutr},
 which,
 after integration with respect to $\mathbf{r}_{||}$ and $z$,
 is
 \begin{equation}\label{electroNeutr2}
  \varrho_0Sl_\slab=eN.
 \end{equation}
 In the case of non-interacting electron gas at low temperatures,
 the number of electrons $N$ has the form
 \begin{equation}\label{electroNeutr3}
   N=\sum_{\mathbf{k}_{||},\alpha}\heav\big(\mathcal{K}_\mathrm{F}^2-\mathbf{k}_{||}^2-\alpha^2\big)
    =\frac{S}{2\pi}\sum_{n=1}^{n_{\max}}\big(\mathcal{K}_\mathrm{F}^2-\alpha_n^2\big),
 \end{equation}
 where two possible orientations of the electron spin are taken into account
 and transition from the sum over two-dimensional vector $\mathbf{k}_{||}$ to the integral is performed according to Refs.~\cite{Kostrobij2015075441,Kostrobij2016155401}.
 We assume that the concentration of the positive charge $\varrho_0$ is equal to the bulk concentration of electrons ${3e/(4\pi r_\mathrm{s}^3)}$,
 where $r_\mathrm{s}$ is the Wigner-Seitz radius.
 Then the condition of electroneutrality becomes
 \begin{equation}\label{electroNeutr4}
   \frac{3}{2r_\mathrm{s}^3}=\frac1{l_\slab}\sum_{n=1}^{n_{\max}}\big(\mathcal{K}_\mathrm{F}^2-\alpha_n^2\big).
 \end{equation}
 This algebraic equation for the magnitude of the Fermi wave vector $\mathcal{K}_\mathrm {F}$,
 which determines the chemical potential ${\mu=\frac{\hbar^2\mathcal{K}_\mathrm{F}^2}{2m}}$ must be solved numerically,
 taking into account the algebraic equation~\eqref{eqForAlfa} for the quantum numbers $\alpha$,
 Eq.~\eqref{nmax} for $n_{\max}$,
 the relation~\eqref{relatForllslab} between the film thickness $l_\slab$,
 the well width $l$,
 and the parameters $d_i$ (${i=1,2}$),
 which, in turn, depend on the unknown magnitude of the Fermi wave vector $\mathcal{K}_\mathrm{F}$.

 \section{Results of numerical calculations and conclusions}

\begin{figure}[htbp]
  \centering
  \includegraphics[width=0.8\textwidth]{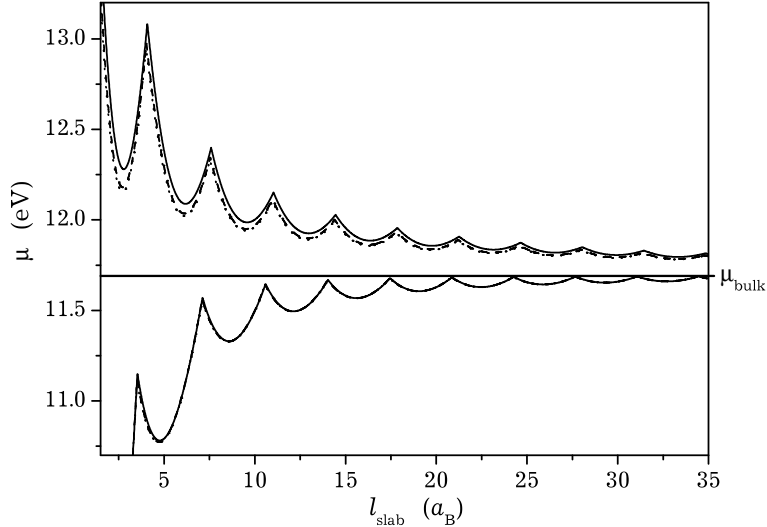}\\
  \caption{The chemical potential $\mu$ as a function of the aluminum film thickness $l_\slab$.
           The upper three curves represent calculations with the incorrect taking into account the electroneutrality condition (i.e. ${l=l_\slab}$)~\cite {Korotun2015391},
           the lower ones --- with the correct taking into account the electroneutrality condition.
           The solid lines are for the aluminium film in the vacuum,
           the dashed lines --- for the aluminium film on SiO$_2$,
           dots ---  for the aluminium film on Al$_2$O$_3$.
           The horizontal solid line show the bulk chemical potential of aluminium within model of non-interacting electrons,
           ${\mu_\bulk=11.695}$\,eV.}\label{mu}
\end{figure}

\begin{figure}[htbp]
  \centering
  \includegraphics[width=0.8\textwidth]{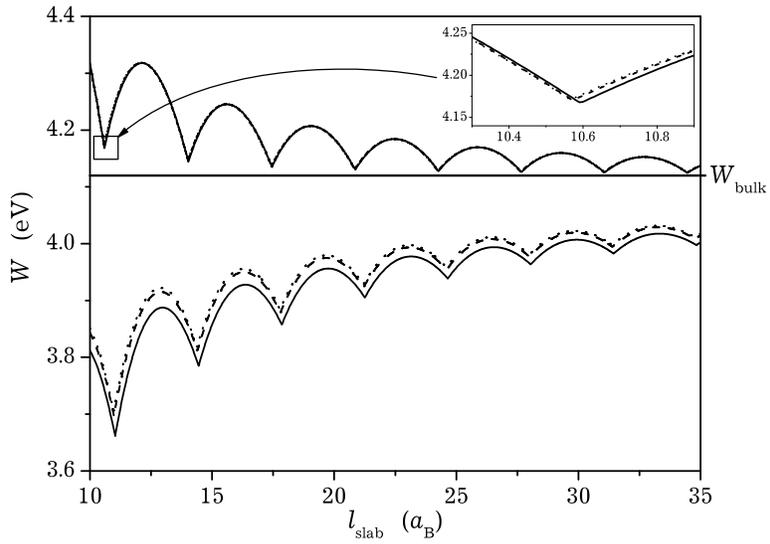}\\
  \caption{The work function $W$ as a function of the aluminum film thickness $l_\slab$.
           The lower three curves represent calculations with the incorrect taking into account the electroneutrality condition (i.e. ${l=l_\slab}$)~\cite {Korotun2015391},
           the upper ones --- with the correct taking into account the electroneutrality condition.
           The solid lines are for the aluminium film in the vacuum,
           the dashed lines --- for the aluminium film on SiO$_2$,
           dots ---  for the aluminium film on Al$_2$O$_3$.
           The horizontal solid line show the bulk work function of aluminium,
           ${W_\bulk=4.119}$\,eV (according to Ref.~\cite{Korotun2015391}).}\label{work}
\end{figure}

 The results of numerical calculations of the chemical potential $\mu$ and the work function ${W=W_1-\mu}$ of electrons for an aluminum film (${r_\mathrm{s}=2.07\,a_\mathrm{B}}$) are given in Fig.~\ref{mu},~\ref{work} for the cases:
 (1) the film is in the vacuum (solid lines) ${W_1=W_2=15.813}$\,eV are the heights of the potential walls,
 (2) the film is on the substrate SiO$_2$ (dashed lines) with the electron affinity $\chi=1.1$\,eV,
 (3)  the film is on the substrate Al$_2$O$_3$ (dots) with the electron affinity $\chi=1.35$\,eV
 (all of these data are taken from Ref.~\cite{Korotun2015391}).

 We see from Fig.~\ref{mu},~\ref{work} that the dependences of the chemical potential and the work function on the film thickness is non monotonic,
 there are alternating peaks,
 i.e. we observe one (oscillatory eﬀect) of two types of the quantum size eﬀect for these characteristics of the metal film.
 This is a consequence of quantization of the electron energy levels,
 because the motion of electrons in the direction perpendicular to the ﬁlm is limited.
 If the film thickness increases,
 the oscillatory effect vanishes.
 As noted by Schulte~\cite{Schulte1976427},
 the distance between adjacent peaks of the chemical potential and the work function is about $\lambda^0_\textrm{F}/2$,
 where $\lambda^0_\textrm{F}=2\pi/\mathcal{K}_\mathrm{F}^0$ is the Fermi wavelength of non-interacting electrons.

 The chemical potential and the work function,
 which are calculated with the correct taking into account the electroneutrality condition,
 approach to their bulk values with increasing of the film thickness.
 In contrast,
 in the case of incorrect taking into account the electroneutrality condition,
 as it was made in Ref.~\cite{Korotun2015391},
 the chemical potential is overestimated (the three upper curves in Fig.~\ref{mu}),
 and the work function is understated (three lower curves in Fig.~\ref{work}).
 This is physically clear,
 since with such an incorrect consideration of the electroneutrality conditions it is assumed that the film thickness is equal to the width of the potential well in which the electrons are situated.
 In fact,
 the film is narrower and quantity of positive charge is less than quantity of  negative charge,
 i.e. the film is negative charged and it is easier for electrons to leave the film, because of this the work function is less than that for the electroneutral film.

 We see from Fig.~\ref{mu} that the presence of the dielectric substrate to the right of the film leads to a decrease in the chemical potential.
 This is physically clear,
 the dielectric substrate reduces the height of the potential barrier ``film--dielectric'',
 and therefore decreases the chemical potential.
 And this leads to an increase in the work function (see Fig.~\ref{work}).
 However,
 the influence of the dielectric substrate with the correct consideration of the electroneutrality condition is less than with incorrect~\cite{Korotun2015391}.
 The reason for this is the small value of the electron affinity of these dielectrics relative to the height of the potential barrier for aluminum in the vacuum.

 Consequently,
 even such the simple model of non-interacting electrons in the field of the asymmetric rectangular potential well with the correct consideration of the electroneutrality condition gives physically reasonable results.
 However,
 for a more precise description of such systems,
 one must also take into account the Coulomb interaction between electrons,
 as was made in Ref.~\cite{Kostrobij2017arxiv},
 and the polarization effects in the dielectric substrate.












\end{document}